%% file: main.tex
\documentclass[11pt]{article}

\usepackage[final]{acl}

\usepackage{times}
\usepackage{latexsym}

\usepackage[T1]{fontenc}

\usepackage[utf8]{inputenc}

\usepackage{microtype}

\usepackage{inconsolata}

\usepackage{graphicx}

\usepackage{booktabs}
\usepackage{enumitem}
\usepackage{array}
\usepackage[table]{xcolor}
\usepackage{tabularx}
\usepackage{makecell}
\usepackage{adjustbox}
\usepackage{multirow}
\usepackage{amssymb}
\usepackage{amsmath}

\usepackage{listings}
\usepackage{pifont} 
\newcommand{\cmark}{\ding{51}} 
\newcommand{\xmark}{\ding{55}} 
\definecolor{oursrow}{HTML}{E0F2FE} 

\definecolor{headergray}{gray}{0.92}
\definecolor{rowgray}{gray}{0.97}

\newcommand{\tablesmall}{\fontsize{8.4}{9.6}\selectfont}
\renewcommand{\arraystretch}{1.08}

\definecolor{tblheader}{HTML}{E2E8F0} 
\definecolor{tblrow}{HTML}{F8FAFC}    
\definecolor{tblblue}{HTML}{DBEAFE}   
\definecolor{textdark}{HTML}{1E293B}  

\usepackage{fancyhdr}

\fancypagestyle{firstpagefooter}{
  \fancyhf{}

  \fancyfoot[C]{%
    \footnotesize
    \begin{tabular}{c}
    Proceedings of the 6th International Conference on Natural Language Processing for Digital Humanities, pages 1--12\\
    July 4, 2026 \copyright2026 Association for Computational Linguistics
    \end{tabular}
  }
}

\title{From OCR to Analysis: Tracking Correction Provenance in Digital Humanities Pipelines}

\author{Haoze Guo\\
  University of Wisconsin - Madison\\
  College of Engineering\\
  \texttt{hguo246@wisc.edu} \\\And
  Ziqi Wei \\
  University of Wisconsin - Madison\\
  College of Letters and Sciences\\
  \texttt{zwei232@wisc.edu} \\}

\begin{document}
\maketitle

\thispagestyle{firstpagefooter}

\begin{abstract}
Optical Character Recognition (OCR) is a critical but error-prone stage in digital humanities text pipelines. While OCR correction improves usability for downstream NLP tasks, common workflows often overwrite intermediate decisions, obscuring how textual transformations affect scholarly interpretation. We present a provenance-aware framework for OCR-corrected humanities corpora that records correction lineage as base-anchored span edits, including edit type, correction source, confidence, and review status. Using a pilot corpus of historical texts, we compare downstream named entity extraction across raw OCR, fully corrected text, and provenance-filtered corrections. Our results show that correction pathways can substantially alter extracted entities and document-level interpretations, while provenance signals help identify unstable outputs and prioritize human review. We argue that provenance should be treated as a first-class analytical layer in NLP for digital humanities, supporting reproducibility, source criticism, and uncertainty-aware interpretation.
\end{abstract}

\section{Introduction}
Optical Character Recognition (OCR) is an important part of the digital humanities (DH) text analysis workflow which allows researchers to build searchable and analyzable corpora from scanned material \citep{smith2007tesseract,nguyen2021postocrsurvey}. OCR is the first step in constructing resource-wide indices and retrieval systems for archival and newspaper collections and provides input for corpus-scale analysis processing pipelines \citep{ehrmann2020impresso,during2021impressoinspect,during2024transparent}. Unfortunately, OCR output produced from historical sources is often noisy because many historical materials are degraded, have non-standard characters, have unusual layouts, and have typographically varied characters \citep{nguyen2021postocrsurvey,neudecker2021ocreval}. Consequently, many researchers use one or more methods to clean up their OCR output prior to performing natural language processing (NLP) such as using normalizations through rules, post-correction through neural networks, and/or manually editing \citep{evershed2014context,nguyen2020bert,lyu2021neural}.

These correction steps improve readability and frequently positively impact downstream NLP performance. However, they can also make it difficult to understand how the original text may have evolved throughout the process and what changes are still uncertain. Corrections can erase analytical history from a text by replacing the original OCR output without any indication of the original content, the nature of the change(s), or which change(s) are uncertain to the analyst. This case represents a methodological challenge for Digital Humanities (DH) because future interpretations of data (e.g., named entities, dates, topics, and trends) will be based on textual transformations that cannot be seen by the analyst \citep{strange2014murder,vanstrien2020impact}. This issue is particularly important for historical named entity recognition (NER) because noisy OCR and historical variation are known to have impacted the quality, reliability, and the ability to evaluate the data extracted from text \citep{hamdi2019analysis,ehrmann2024survey,boros2020alleviating}.

We suggest that OCR correction be not just one way; but should be modeled as a repeated sequence of traceable editorial decisions; the provenance of which are also able to be examined with downstream NLP outputs. In the DH context this framing is especially important since core DH values \citep{ockeloen2013biographynet,romein2024htrprov} include source criticism, uncertainty and interpretive transparency. Previous DH infrastructure work \citep{ockeloen2013biographynet, guo2026feedtaxonomyuserfacingcues} has already emphasised the need for multi-level provenance as a significant requirement of scholarly workflows; therefore, broader provenance standards provide a formal representation of such traces \citep{moreau2013provdm,lebo2013provo,moreau2013introprov}. This context can also be seen by looking at broader audit-focused research that discusses the need for preserving transformation traces for longitudinal analysis and assessment \citep{guo2026temporaldriftprivacyrecall,10.1145/3772363.3798570,10.1145/3774904.3792853}, even across application domains.

In this paper, we introduce a provenance-aware representation for OCR-corrected humanities corpora and demonstrate its utility in a pilot study. We make three contributions:
\begin{itemize}[leftmargin=*, itemsep=2pt]
    \item We introduce a span-level provenance schema for OCR correction that records edit lineage, correction source, confidence, and review status.
    \item We present a pilot empirical comparison of downstream named entity recognition (NER) across raw OCR, fully corrected text, and provenance-filtered corrected text.
    \item We provide a DH-oriented error analysis lens showing how provenance signals can identify unstable outputs and prioritize human review.
\end{itemize}

\section{Related Work}
Historical OCR pipelines face challenges beyond standard contemporary OCR settings. Noise can arise from page damage, bleed-through, historical fonts, line segmentation errors, footnotes, marginalia, and spelling conventions that differ from modern forms \citep{nguyen2021postocrsurvey,neudecker2021ocreval}. As a result, OCR errors are not uniformly distributed across documents or regions of a page, and ``correction'' may involve both error repair and editorial normalization \citep{evershed2014context,nguyen2021postocrsurvey}.

For DH scholars, this distinction matters. A correction that restores a malformed token may improve fidelity to the source, while a normalization step may improve computational consistency but alter historically meaningful variation \citep{strange2014murder}. In both cases, downstream NLP outputs can shift: entities may appear or disappear, names may merge or split, and frequencies may change \citep{vanstrien2020impact,hamdi2019analysis,ehrmann2024survey}. If the intermediate correction decisions are not preserved, it becomes difficult to audit why a downstream result changed.

This issue is especially salient for named entity analysis on historical texts. Shared tasks and survey work (e.g., HIPE) document the difficulty of robust NER and linking in multilingual historical materials, including challenges from noisy OCR, diachronic variation, and annotation heterogeneity \citep{ehrmann2020clefhipe,ehrmann2022hipe,ehrmann2024survey}. Parallel work on historical newspaper infrastructures such as \textit{impresso} has also emphasized the need to connect computational processing with scholarly interpretability and interface-level transparency \citep{ehrmann2020impresso,during2021impressoinspect,during2024transparent}.

Our motivation is therefore not only to improve OCR correction quality, but to preserve and expose the lineage of correction decisions so that computational findings remain inspectable and reproducible. Provenance research offers a strong conceptual foundation for this goal, including explicit modeling of entities, activities, and agents \citep{moreau2013provdm,lebo2013provo}, but DH OCR pipelines need a lightweight, text-first representation that can operate at the span level and integrate directly with downstream NLP outputs. 

We summarize how common provenance/annotation and OCR-output practices relate to the requirements of replayable span edits and policy-driven variant construction in Table~\ref{tab:comparison-prior}. This comparison motivates our focus on base-anchored span-edit events coupled with explicit trust policies for reconstruction and audit.

\begin{table*}[t]
  \centering
  \resizebox{\textwidth}{!}{%
    \rowcolors{2}{white}{gray!10}%
    \begin{tabular}{>{\bfseries}l c c c c c c}
      Approach
        & Span-level edits
        & Base-anchored offsets
        & Deterministic replay
        & Trust-policy filtering
        & Downstream trace links
        & Pipeline-ready artifact \\
      \toprule
      PROV-DM / PROV-O \citep{moreau2013provdm,lebo2013provo}
      & \xmark & \xmark & \xmark & \xmark & \cmark & \xmark \\
    TEI stand-off (\texttt{<standOff>}) \citep{tei_standoff}
      & \cmark & \cmark & \xmark & \xmark & \xmark & \cmark \\
    ALTO XML \citep{alto_loc}
      & \xmark & \cmark & \xmark & \xmark & \xmark & \cmark \\
    hOCR \citep{hocr_spec}
      & \xmark & \cmark & \xmark & \xmark & \xmark & \cmark \\
    PAGE-XML \citep{pletschacher2010page}
      & \xmark & \cmark & \xmark & \xmark & \xmark & \cmark \\
    OCR-D \citep{neudecker2019ocrd,ocrd_mets,ocrd_workflows}
      & \xmark & \cmark & \xmark & \xmark & \xmark & \cmark \\
    \rowcolor{oursrow}
    Span-edit events + provenance-aware filtering (Ours)
      & \cmark & \cmark & \cmark & \cmark & \cmark & \cmark \\
      \bottomrule
    \end{tabular}
  }
  \caption{Comparison of our span-edit provenance + policy-driven reconstruction with common DH/OCR provenance and annotation practices.}
  \label{tab:comparison-prior}
\end{table*}

\section{Provenance-Aware Correction Schema}
\subsection{Design Principles}
We design the schema around three principles: (1) \textbf{traceability}, so each correction can be linked to a document location and prior text span; (2) \textbf{tool-agnostic interoperability}, so the representation can capture rule-based, model-based, and human edits; and (3) \textbf{analytical usefulness}, so metadata can support downstream filtering and audit. Our design is informed by both DH provenance modeling practices and the broader PROV family of standards \citep{ockeloen2013biographynet,moreau2013provdm,lebo2013provo}.

We represent corrections at the \textbf{span level}, which provides more flexibility than token-only records while remaining simpler than character-level edit traces. A span may correspond to a token, multi-token phrase, or split/merge operation. This design is compatible with OCR post-correction workflows ranging from noisy-channel approaches to neural sequence models \citep{evershed2014context,nguyen2020bert,lyu2021neural,nguyen2021postocrsurvey}.

\subsection{Schema}
We represent OCR correction provenance at the span level using a compact record schema that links each edit to a document location and revision step. Each record stores: document/page identifiers, span offsets, original and corrected text, edit type (e.g., substitution, split, merge), correction source (rule-based, model-assisted, or human), optional confidence, optional review status, and optional layout-zone metadata (e.g., body, header, footnote). This representation is sufficient to reconstruct correction lineage and to trace downstream NLP outputs (e.g., extracted entities) back to the edits they depend on \citep{ciccarese2013pav}.

\subsection{Illustrative Record}
A single provenance record links an OCR span to its corrected form together with the metadata needed for audit (e.g., edit type, correction source, confidence, and review status). For example, if the OCR span \textit{Madifon} is corrected to \textit{Madison} by a model-assisted step with confidence 0.74 and no human approval, any downstream entity extracted from \textit{Madison} can be traced back to that specific correction event. This span-level linkage is central to our analysis of entity volatility and uncertainty-aware filtering in the pilot study.

\subsection{Serialization and Interoperability}
Our span-level provenance schema is designed to be tool-agnostic and easy to integrate into existing NLP pipelines. The correction records can be serialized as \textbf{JSONL}, as a \textbf{tabular file} (CSV/Parquet), or as \textbf{stand-off annotations} that reference the underlying text by character offsets. Stand-off serialization is particularly useful for DH settings because it preserves the original OCR text while enabling multiple correction layers to be applied or compared without overwriting intermediate states.

In JSON-based NLP pipelines, provenance records can be embedded as an auxiliary field alongside the raw and corrected text variants, enabling downstream tasks to (i) reconstruct a chosen variant under a specified trust policy, and (ii) trace extracted entities back to the correction events that influenced them. For interoperability across tools, we recommend storing: stable document/page identifiers, explicit span offsets, explicit revision steps, and correction-source metadata, so that records remain portable even when tokenization or sentence segmentation changes. This makes the schema compatible with common DH workflows and with standard NLP processing stacks that expect JSON-serializable artifacts.

\subsection{Offset Semantics and Edit Application}
All correction events are anchored to a \textbf{base revision} (typically raw OCR) to avoid cascading offset drift and to preserve stable linkage from downstream outputs back to scan-derived text. Offsets (\texttt{span\_start}, \texttt{span\_end}) are Unicode-codepoint indices over the base text using half-open intervals \texttt{[start,end)}; \texttt{orig\_text} must exactly match the base substring at that interval as an integrity check. To construct a variant, we select events under a trust policy (e.g., confidence threshold or \texttt{review\_status=approved}), sort by \texttt{span\_start} (tie-break by \texttt{event\_id}), and apply them as base-anchored replacements, explicitly detecting overlaps for resolution (e.g., prefer human over model or defer to adjudication). Boundary-sensitive edits (split/merge) and OCR normalizations (line-break/whitespace reflow, hyphenation repair such as \texttt{"inter-\\national"}$\rightarrow$\texttt{"international"}, paragraph merges) are represented as the same span-replacement events with explicit boundaries; large reflow regions are recorded as a single replacement to avoid brittle micro-edits.

\section{Pilot Study Design}

\subsection{Corpus and Unit of Analysis}
We conduct a pilot study on a small corpus of historical texts drawn from a digitized humanities collection. The corpus includes scanned materials with OCR output and a subset of passages that undergo correction. We treat this study as a methodological pilot rather than a benchmark-scale evaluation, with the goal of testing whether provenance signals provide useful analytical leverage in downstream NLP \citep{vanstrien2020impact}.

Our primary unit of analysis is the \textbf{document-level NER output} for each text variant, with additional linkage to \textbf{span-level correction records}. For each document, we preserve the raw OCR text and record every applied correction as a provenance event, including span offsets, edit type, correction source, and optional confidence/review status. This design supports both aggregate comparison, how the extracted entity inventory changes across variants, and auditability which correction events are associated with unstable entities.

\subsection{Text Variants and Provenance Filtering Policy}
For each document, we construct three text variants from the same OCR source: \textbf{Raw OCR} (no post-correction), \textbf{Fully corrected} (all available corrections applied), and \textbf{Provenance-filtered} (only corrections meeting a provenance criterion).

The provenance-filtered variant operationalizes a conservative ``trust policy'' over corrections. In the main comparison, this policy is confidence-based (e.g., confidence $\geq 0.70$) and may optionally require human approval when such metadata is available. We additionally report a threshold sensitivity analysis (Table~\ref{tab:threshold-sensitivity}) that varies the filter strictness, including higher confidence thresholds and a human-approved-only condition. This provides a transparent characterization of the coverage--stability tradeoff induced by provenance-aware filtering, rather than presenting a single corrected corpus as an implicit ground truth.

This design separates two pipeline questions that are often conflated in DH OCR workflows: (i) whether correction improves downstream extraction at all, and (ii) which corrections should be treated as sufficiently reliable for interpretive analysis.

\subsubsection{Where Confidence Comes From}
In our pilot, \texttt{confidence} is an optional provenance attribute attached to each correction event and emitted by the correction pathway that proposed the edit. For \textbf{model-assisted} corrections, \texttt{confidence} corresponds to the model's own edit score, used to rank suggested substitutions and split/merge operations. For \textbf{rule-based} corrections, \texttt{confidence} is derived from rule certainty (e.g., high-confidence deterministic patterns versus lower-confidence heuristic matches). For \textbf{human} corrections, we treat \texttt{review\_status=approved} as the primary reliability signal; when an explicit reviewer score is available, we store it as \texttt{confidence}.

These scores are not assumed to be calibrated probabilities, and they are not assumed comparable across different tools or correction sources. Accordingly, we use \texttt{confidence} only as a \emph{within-pipeline ranking signal} for sensitivity analysis. Thresholds (e.g., confidence $\geq 0.70$ in the main provenance-filtered condition) are applied under a fixed correction stack and interpreted operationally rather than as a universal notion of correctness. The threshold sweep (Table~\ref{tab:threshold-sensitivity}) is therefore intended to make the coverage--stability tradeoff explicit under a fixed scoring regime, not to claim that a given numeric threshold generalizes across tool stacks.

\subsection{Downstream Task and Quantitative Metrics}
We use named entity recognition (NER) as the downstream task because entity extraction is common in DH workflows and is highly sensitive to OCR noise, normalization, and segmentation decisions \citep{hamdi2019analysis,boros2020alleviating,ehrmann2020clefhipe,ehrmann2022hipe,ehrmann2024survey,clausner2011scenario}. We run the same NER pipeline over all text variants and compare the resulting entity outputs.

\subsubsection{NER Pipeline Configuration}
We use a single, fixed NER pipeline across all text variants to isolate the effect of OCR correction and provenance filtering. Concretely, we sentence-segment the text, tokenize using the model's native subword tokenizer, and apply a transformer-based NER model fine-tuned on CoNLL-2003 (implemented via the HuggingFace Transformers library) \citep{devlin2019bert,liu2019roberta,wolf2020transformers}. We keep all inference settings fixed across variants.

Because historical corpora exhibit diachronic spelling and orthographic variation, we avoid additional spelling modernization or normalization inside the NER pipeline beyond what is already implied by the OCR correction variants under study. This design prevents conflating ``downstream robustness tricks'' with the provenance-aware correction analysis. We acknowledge that domain adaptation can substantially affect absolute NER accuracy in historical settings; our goal here is to characterize how correction pathways change extracted entities under a consistent downstream model \citep{ehrmann2024survey,hamdi2019analysis,boros2020alleviating}.

To quantify differences across variants, we report:
\begin{itemize}[leftmargin=*, itemsep=2pt]
    \item \textbf{Entity mentions}: total number of extracted entity mentions per variant.
    \item \textbf{Unique entities}: number of distinct extracted entity surface strings per variant.
    \item \textbf{Entity overlap}: Jaccard similarity between unique-entity sets across variants.
    \item \textbf{Entity volatility}: entities that (i) appear in one variant but not another, or (ii) appear in both but with changed surface form or span boundary.
\end{itemize}

In addition, we compute the share of volatile entities linked to \textbf{low-confidence or unreviewed correction spans}. This measure provides an audit-oriented view of instability: instead of treating volatility as an opaque downstream artifact, it tests whether provenance fields can identify subsets of edits that disproportionately contribute to unstable outputs.

\subsection{Entity Linking as a Secondary Downstream Task}
To test whether provenance-aware correction affects not only entity extraction but also downstream \emph{resolution} of entities to knowledge bases, we include a lightweight entity linking (EL) analysis on a pilot subset. For each text variant, we take extracted entity mentions and apply an off-the-shelf EL system to link mentions to a reference knowledge base (e.g., Wikipedia/Wikidata identifiers). We focus on mentions that are volatile across variants, since these are the cases where small surface-form or boundary changes are most likely to alter linking decisions.

We report (i) \textbf{linking coverage} (fraction of mentions assigned a KB identifier), (ii) \textbf{link stability} across variants, fraction of mentions that resolve to the same KB identifier, and (iii) a small \textbf{manual audit of disagreements} to characterize whether changes reflect improved disambiguation, harmful normalization, or overconfident linking on noisy strings. This EL check is designed as a methodological stress test rather than a benchmark-scale evaluation.

\subsection{Attribution of Entity Changes to Correction Events}
To associate downstream entity differences with provenance records, we compute an \textbf{association} (not causal proof) between each entity mention and nearby correction events. We first attempt \textbf{span overlap}: an entity mention is associated with a correction record if the mention character offsets overlap the correction span offsets in that same variant. For edits that may shift offsets (notably split/merge operations), we use a bounded local fallback: we search for correction events within a window of $\pm W$ characters around the entity mention (we set $W=50$ in the pilot) and choose the closest event by absolute offset distance. Ties are broken by preferring (i) overlap over proximity, (ii) edit types that can directly affect boundaries (split/merge) over substitutions, and (iii) higher-confidence events when available.

We use this association to support audit and summarization (e.g., ``how many volatile entities are linked to low-confidence edits''), and we interpret results as identifying \emph{likely contributing edits} rather than definitive causes.

\subsection{Validation of the Attribution Heuristic}
Because proximity-based association can produce spurious links (nearby edits may be unrelated), we validate the attribution heuristic on a labeled sample. We randomly sample linked (entity, correction-event) pairs and ask annotators to judge whether the correction plausibly contributed to the entity change (yes/no). We report the precision of the heuristic and use this validation to qualify interpretation of the signal-utility analysis.

\subsection{Qualitative Coding Protocol}
To complement the quantitative comparison, we manually inspect a pilot sample of volatile entities and assign each case to an error category (e.g., OCR noise, normalization shift, split/merge boundary issue, layout artifact). The goal of this coding step is not to produce an exhaustive taxonomy, but to evaluate whether provenance fields (confidence, edit type, layout zone, review status) provide actionable diagnostic value. We use the qualitative categories to interpret the mechanisms behind volatility and to assess which provenance signals are most useful for DH-oriented review and source criticism.

\section{Results}

\subsection{Quantitative Differences Across Correction Pathways}
Across the pilot corpus, correction activity is unevenly distributed across documents and page regions, and many edits occur in short spans that affect candidate entity strings. This matters because a relatively small number of span edits can produce disproportionately large downstream changes in NER outputs, particularly when edits touch capitalization, rare proper nouns, or token boundaries.

\begin{figure}[t]
    \centering
    \includegraphics[width=0.8\columnwidth]{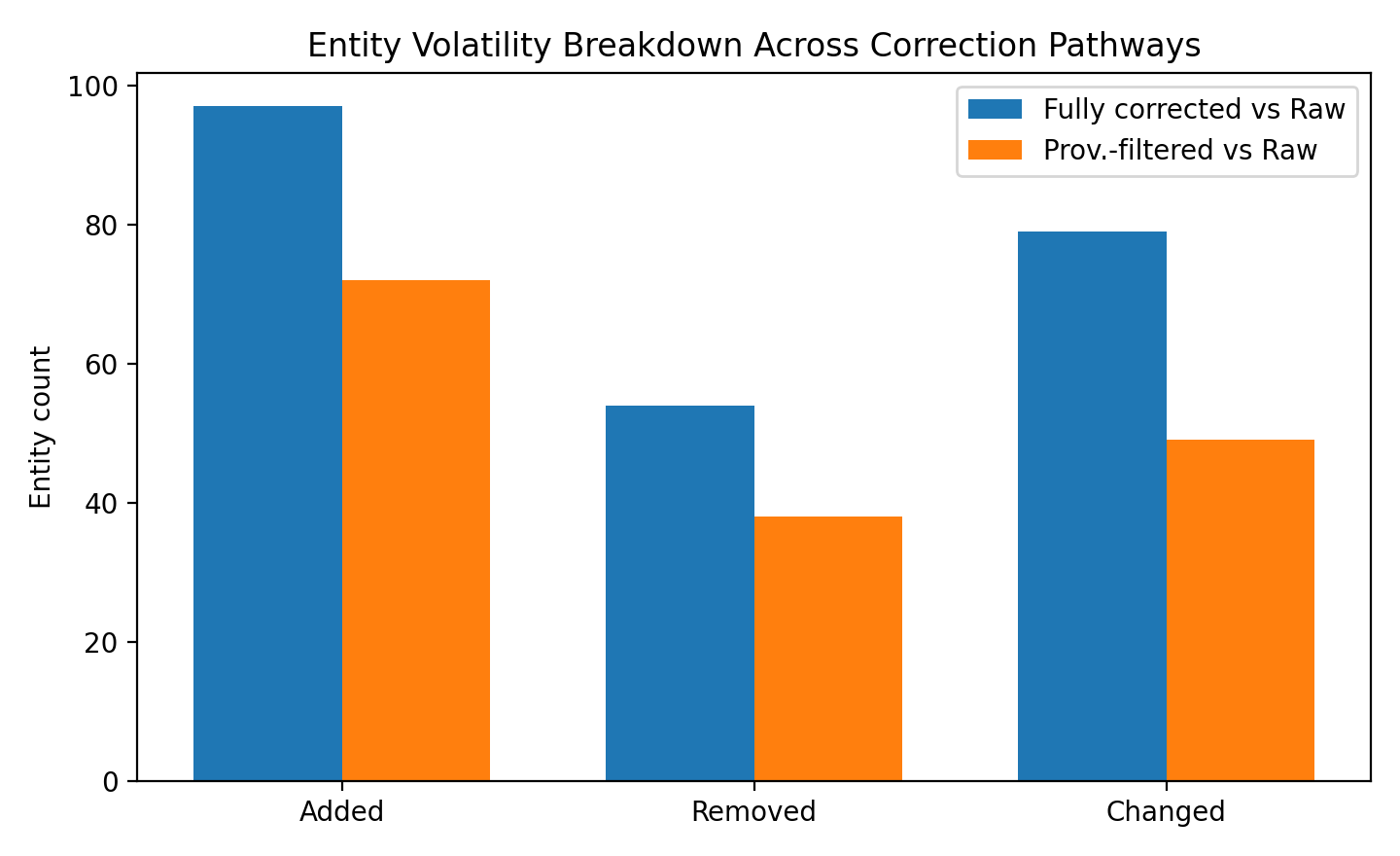}
    \caption{Breakdown of entity volatility across correction pathways.}
    \label{fig:volatility-breakdown}
\end{figure}

Figure~\ref{fig:volatility-breakdown} provides a compact view of entity volatility relative to raw OCR. The fully corrected condition produces more \emph{added} and \emph{changed} entities than the provenance-filtered condition, indicating stronger transformation of the extracted entity inventory. The provenance-filtered condition still preserves a substantial portion of correction gains, but reduces high-risk changes. In practice, this suggests that provenance filtering behaves less like ``undoing correction'' and more like selecting a correction pathway with a different analytical risk profile.

Table~\ref{tab:main-results} quantifies these differences. Relative to raw OCR, the fully corrected variant increases both entity mentions (1184 $\rightarrow$ 1342) and unique entities (512 $\rightarrow$ 566), consistent with correction improving recognizability of names and places. However, these gains coincide with the highest volatility (176 volatile entities), meaning that correction alters not only the volume of extracted entities but also the stability of the extracted inventory. The provenance-filtered variant retains most of the coverage gains (1287 mentions; 548 unique entities) while reducing volatility (121), suggesting that provenance-aware filtering can improve analytical stability without reverting to raw OCR.

A key observation is that volatility is not uniformly distributed across edits: a substantial fraction of volatile entities in the corrected conditions is linked to low-confidence or unreviewed correction events (Table~\ref{tab:main-results}). This does not imply that all low-confidence edits are incorrect; rather, it indicates that these edits are most likely to produce downstream differences that analysts may want to inspect, qualify, or report explicitly.

\begin{table}[t]
\centering
\tablesmall
\setlength{\tabcolsep}{3pt}
\renewcommand{\arraystretch}{1.05}
\begin{tabularx}{\columnwidth}{l r r c r c}
\toprule
\rowcolor{tblheader}
\textbf{Variant} & \textbf{Ment.} & \textbf{Uniq.} & \textbf{Jac. vs Raw} & \textbf{Vol.} & \textbf{\% Unrev.} \\
\midrule
\rowcolor{tblrow}
Raw OCR         & 1184 & 512 & 1.00 & --  & -- \\
Fully corrected & 1342 & 566 & 0.69 & 176 & 68\% \\
\rowcolor{tblrow}
Prov.-filtered  & 1287 & 548 & 0.76 & 121 & 76\% \\
\bottomrule
\end{tabularx}
\caption{NER output differences across OCR correction pathways. ``Vol.'' counts entities that appear/disappear or change surface form/span boundary across variants; ``\% Unrev.'' is the share of volatile entities linked to unreviewed edits.}
\label{tab:main-results}
\end{table}

\subsection{Threshold Sensitivity of Provenance Filtering}
Table~\ref{tab:threshold-sensitivity} shows how downstream outputs change as the provenance filter becomes stricter. As the confidence threshold increases (or when only human-approved edits are used), entity volatility decreases monotonically, but entity coverage also declines. This tradeoff is expected and analytically useful: different DH workflows may prefer different operating points depending on whether the priority is recall-oriented exploration or conservative interpretation.

\begin{figure}[t]
    \centering
    \includegraphics[width=0.9\columnwidth]{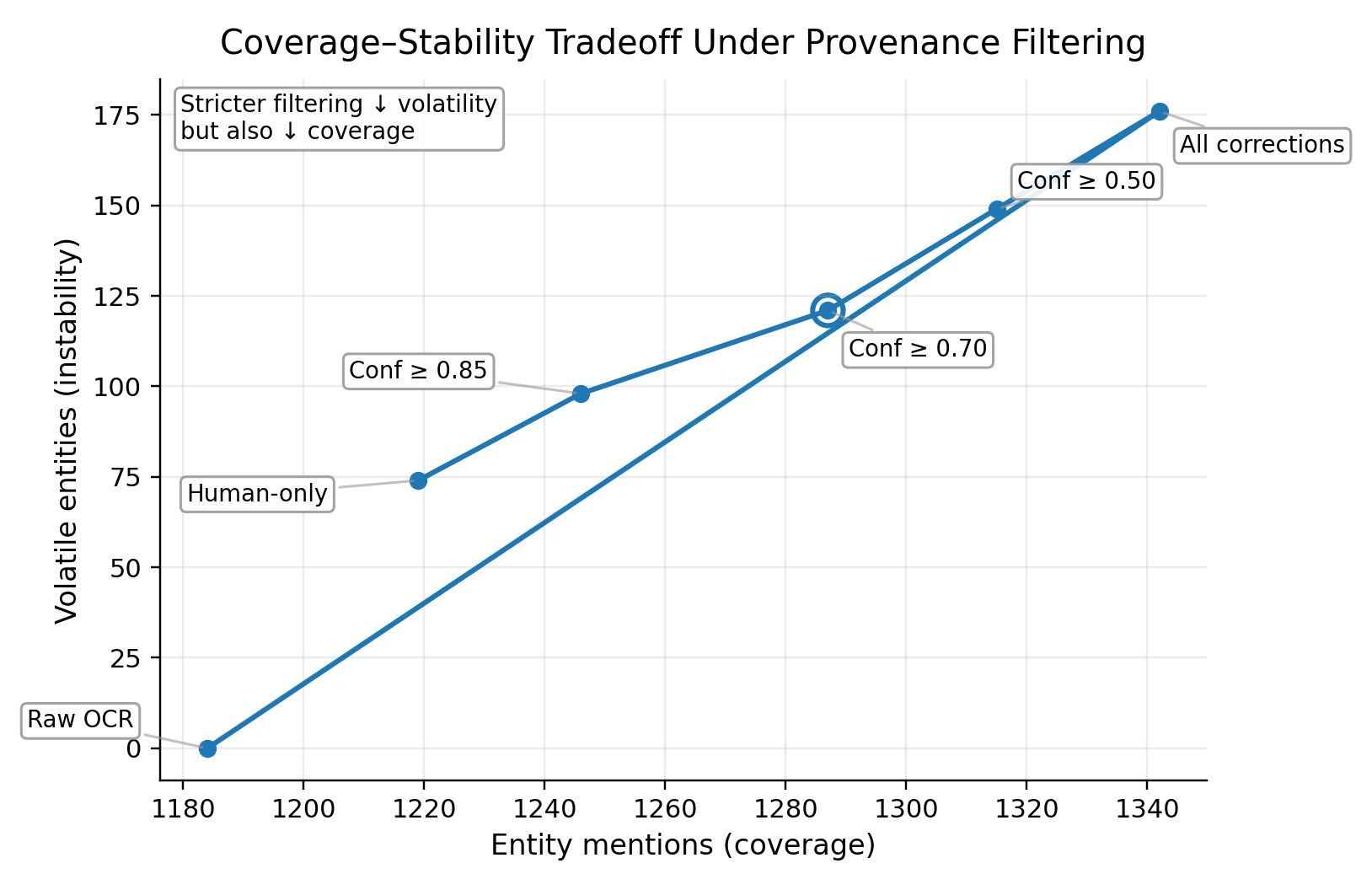}
    \caption{Coverage--stability tradeoff under provenance filtering.}
    \label{fig:tradeoff}
\end{figure}

Figure~\ref{fig:tradeoff} visualizes the same sweep as a coverage--stability curve. Moving from ``All corrections'' to stricter thresholds shifts the operating point toward lower instability, but also lower coverage. In this pilot, a mid-range threshold (e.g., confidence $\geq 0.70$) yields a balanced regime: many of the coverage improvements from correction are retained, while a substantial portion of volatility is reduced. We emphasize that thresholds are not universal: confidence values are tool- and workflow-dependent. The contribution here is that provenance makes the operating point explicit and reportable, rather than implicit in a single overwritten corrected text.

\begin{table}[t]
\centering
\tablesmall
\setlength{\tabcolsep}{3pt}
\renewcommand{\arraystretch}{1.05}
\begin{tabular}{lrrrr}
\toprule
\rowcolor{headergray}
\textbf{Filter} & \textbf{Mentions} & \textbf{Unique} & \makecell{\textbf{Jaccard}\\\textbf{vs Raw}} & \textbf{Volatile} \\
\midrule
\rowcolor{rowgray}
Raw OCR              & 1184 & 512 & 1.00 & -- \\
All corrections      & 1342 & 566 & 0.69 & 176 \\
\rowcolor{rowgray}
Conf $\geq$ 0.50     & 1315 & 559 & 0.73 & 149 \\
Conf $\geq$ 0.70     & 1287 & 548 & 0.76 & 121 \\
\rowcolor{rowgray}
Conf $\geq$ 0.85     & 1246 & 531 & 0.79 & 98 \\
Human-approved only  & 1219 & 523 & 0.82 & 74 \\
\bottomrule
\end{tabular}
\caption{Threshold sensitivity for provenance-filtered correction.}
\label{tab:threshold-sensitivity}
\end{table}

\subsection{Provenance Signals Predicting Instability}
Beyond overall thresholds, provenance fields help identify which correction operations are most likely to affect downstream entity extraction. Table~\ref{tab:signal-utility} summarizes the utility of several provenance signals for predicting volatility. Two patterns are particularly salient. First, boundary-affecting edits (split/merge) exhibit the highest volatility lift despite being relatively infrequent, consistent with entity extraction being sensitive to segmentation decisions. Second, non-body layout zones (e.g., headers and footnotes) are instability hotspots, reflecting known OCR challenges in regions where typography and layout differ from running prose.

Low confidence and unreviewed status provide coarse but useful uncertainty flags. These fields are less diagnostic than edit type or layout zone, but they are broadly applicable across correction pathways and can support conservative analysis regimes (e.g., confidence-based filtering) and targeted inspection when human review capacity is limited.

\begin{table}[t]
\centering
\small
\setlength{\tabcolsep}{3pt}
\renewcommand{\arraystretch}{1.05}
\begin{tabularx}{\columnwidth}{>{\raggedright\arraybackslash\bfseries}X c c c}
\toprule
\rowcolor{tblheader}
\textbf{Signal} & \textbf{\%} & \textbf{Flag vol.} & \textbf{Lift} \\
\midrule
\rowcolor{tblrow}
Low conf.\ $(<0.70)$ & 31\% & 7.8\% & 2.7$\times$ \\
Unreviewed & 90\% & 5.9\% & 1.7$\times$ \\
\rowcolor{tblrow}
Split/merge & 9\% & 12.5\% & 3.3$\times$ \\
Non-body zone & 14\% & 10.1\% & 2.6$\times$ \\
\bottomrule
\end{tabularx}
\caption{Utility of provenance signals for identifying instability. ``Flag vol.'' is the volatility rate among entities associated with the flagged condition; ``Lift'' compares flagged vs unflagged.}
\label{tab:signal-utility}
\end{table}

These associations are produced by a deterministic overlap-plus-window heuristic and should be interpreted as likely contributing edits rather than definitive causal attributions; we therefore validate heuristic precision on a labeled sample and report it alongside the signal-utility analysis.

\subsection{Qualitative Error Categories and Diagnostic Use}
Table~\ref{tab:error-categories} summarizes qualitative categories observed among volatile entities. The most common category is false entities caused by OCR noise, followed by normalization shifts and boundary errors (split/merge cases). These patterns align with known OCR and historical NER challenges \citep{vanstrien2020impact,ehrmann2024survey}, but the key result in our setting is that provenance metadata helps localize and explain failures rather than only reporting aggregate quality.

Different provenance fields are useful for different diagnostic purposes. Confidence and review status help identify risky corrections for triage; edit type helps distinguish normalization shifts from segmentation effects; and layout-zone metadata supports page-region-based auditing of headers, footnotes, and other non-body text. Taken together, provenance enables analysts to answer not only ``what changed'' (entity volatility), but also ``which editorial operations caused the change'' and ``which changes are plausible candidates for targeted review.''

\subsection{Entity Linking Sensitivity to Correction Pathways}
Entity linking is particularly sensitive to surface form and boundary decisions: small spelling changes, whitespace insertion, or normalization can alter candidate generation and disambiguation. In our pilot EL analysis, fully corrected text increases linking coverage relative to raw OCR by producing more linkable surface strings, but it also introduces additional link instability for a subset of volatile mentions. Provenance-filtered text typically preserves much of the coverage gain while reducing the rate of link changes tied to low-confidence or unreviewed corrections.

Qualitatively, we observe three recurring EL disagreement patterns: (i) \textbf{repair-driven improvements} where correcting OCR noise enables linking to the intended entity, (ii) \textbf{normalization-driven shifts} where historical spellings or abbreviations are modernized and shift candidate selection, and (iii) \textbf{boundary-driven failures} where split/merge edits change mention spans and trigger different candidate sets. These results reinforce the core claim that correction pathways are analytically consequential: downstream interpretations can shift not only in \emph{which} entities are extracted, but also in \emph{which real-world entities} those mentions are resolved to.

\begin{table}[t]
\centering
\tablesmall
\setlength{\tabcolsep}{3pt}
\renewcommand{\arraystretch}{1.05}
\begin{tabularx}{\columnwidth}{>{\raggedright\arraybackslash\bfseries}p{2.05cm} c X}
\toprule
\rowcolor{tblheader}
\textbf{Pattern} & \textbf{\%} & \textbf{Diagnostic provenance cue(s)} \\
\midrule
\rowcolor{tblrow}
OCR-noise false entity & 29\% & Low confidence; model-assisted; often disappears under filtering \\
Normalization shift & 21\% & \texttt{edit\_type=substitution}; unreviewed; surface-form changes \\
\rowcolor{tblrow}
Boundary (split/merge) & 18\% & \texttt{edit\_type=split/merge}; span boundary drift \\
Layout artifact & 17\% & \texttt{layout\_zone=header/footnote}; non-body text \\
\rowcolor{tblrow}
Ambiguous correction & 15\% & Medium confidence; unreviewed; competing entity strings \\
\bottomrule
\end{tabularx}
\caption{Qualitative categories among volatile entities.}
\label{tab:error-categories}
\end{table}

\section{Discussion}
The pilot study shows that OCR correction pathways can produce meaningfully different analytical corpora from the same scanned source. Raw OCR, fully corrected text, and provenance-filtered text do not merely differ in readability; they produce different entity inventories, different levels of volatility, and different evidentiary risks for downstream interpretation. This supports our central claim that OCR correction should be treated as an inspectable analytical process rather than a hidden preprocessing step.

Provenance makes these differences explicit. By preserving edit source, confidence, review status, edit type, and layout-zone information, analysts can treat correction as a set of documented editorial choices rather than as an invisible transformation. The threshold sweep and coverage--stability curve show how different trust policies expose different operating points: a recall-oriented analysis may prefer broader correction coverage, while a conservative interpretive analysis may prefer provenance-filtered text with lower volatility.

The qualitative categories highlight a recurring DH concern: some corrections restore fidelity to the source (repair), while others normalize historically meaningful variation (normalization). Provenance makes these transformations visible and therefore contestable. This is important for interpretive work where spelling variation, abbreviations, and typography may carry historical meaning.

Provenance also supports locating where instability originates. Our signal-utility analysis suggests that boundary edits and non-body layout zones are particularly consequential. This enables a more DH-aligned auditing practice in which analysts can attribute changes in extracted entities to specific editorial operations, rather than treating entity differences as opaque model behavior.

\subsection{Limitations and Future Work}
This study has several limitations. First, it is a methodological pilot rather than a benchmark-scale evaluation, and the corpus is not intended to represent the full diversity of DH materials. Future work should test the schema across multiple DH domains, including historical newspapers, literary prose, poetry, correspondence, and archival administrative records, because each genre introduces different OCR and correction failure modes. Second, we use NER as the primary downstream probe; additional tasks such as topic modeling, text classification, relation extraction, and entity linking at larger scale may expose different kinds of correction sensitivity. Additionally, although confidence scores and metadata can be utilized with some degree of familiarity to compare various corrections, they serve as a signal of provenance rather than as calibrated probabilities.

Future research into the analysis of correction pathways which account for provenance must be extended to cover multilingual collections and more complex historical page layouts. In these instances, greater differences in typography and segmentation between different collections will be present. A second area for further research is to expand the downstream tasks used for example with regard to entity linking where small changes in the surface (visibility) features result in significant impact on linking decisions. Lastly, the construction of provenance-aware pipelines creates an opportunity to establish standardized reporting criteria for DH-related NLP. Researchers will be able to use these documents as a means of reporting their own sensitivity to correction pathways while also documenting the provenance signals that directly contributed most to their instability.

\section{Conclusion}
We presented a provenance-aware framework for OCR-corrected humanities corpora and a pilot study showing how correction pathways alter downstream named entity extraction. By preserving correction lineage at the span level, our approach helps make NLP outputs more auditable, reproducible, and interpretable in DH workflows. We argue that provenance should be treated as a first-class analytical layer in OCR-to-NLP pipelines, not merely as implementation metadata.

\bibliography{refs}

\appendix
\input{Appendix}

\end{document}

%% file: Appendix.tex
\section{Schema Specification and Serialization}
\label{app:schema-serialization}

\subsection{Normative field specification}
We represent OCR correction provenance as a sequence of \textbf{span-edit events} anchored to a \textbf{base revision} of the text (typically raw OCR). Each event specifies \emph{where} an edit applies in the base text, \emph{what} replacement occurs, and \emph{metadata} that supports audit and policy-based reconstruction.

\paragraph{Required invariants.}
Each event MUST satisfy \emph{base anchoring} (offsets refer to \texttt{base\_revision}, with \texttt{base\_revision=0} for raw OCR in this paper), \emph{half-open spans} (\texttt{[span\_start, span\_end)} with \texttt{span\_start < span\_end}), \emph{Unicode-codepoint offsets}, and an \emph{integrity check} (\texttt{orig\_text} matches the base substring at \texttt{[span\_start, span\_end)}).

\paragraph{Event fields.}
Table~\ref{tab:app-schema-fields} defines the canonical fields and constraints used by the schema described in \S3 and referenced throughout the pilot study.

\begin{table}[t]
\centering
\tablesmall
\setlength{\tabcolsep}{3pt}
\renewcommand{\arraystretch}{1.05}
\begin{tabularx}{\columnwidth}{l l c X}
\toprule
\rowcolor{tblheader}
\textbf{Field} & \textbf{Type} & \textbf{Req.} & \textbf{Meaning / constraints} \\
\midrule
\rowcolor{tblrow}
\texttt{schema\_version} & string & Y & Semantic version, e.g., \texttt{"1.0.0"}. \\
\texttt{event\_id} & string & Y & Unique identifier within a corpus (UUID recommended). \\
\rowcolor{tblrow}
\texttt{doc\_id} & string & Y & Stable document identifier. \\
\texttt{page\_id} & string/int & Y & Page identifier (or region id if pageless). \\
\rowcolor{tblrow}
\texttt{base\_revision} & int & Y & Revision index of anchor text (0 = raw OCR). \\
\texttt{span\_start} & int & Y & Inclusive offset in base revision (Unicode codepoints). \\
\rowcolor{tblrow}
\texttt{span\_end} & int & Y & Exclusive offset; \texttt{span\_start < span\_end}. \\
\texttt{orig\_text} & string & Y & Exact base substring at \texttt{[span\_start, span\_end)}. \\
\rowcolor{tblrow}
\texttt{new\_text} & string & Y & Replacement string; empty string indicates deletion. \\
\texttt{edit\_type} & enum & Y & \{\texttt{substitute}, \texttt{insert}, \texttt{delete}, \texttt{split}, \texttt{merge}, \texttt{normalize}\}. \\
\rowcolor{tblrow}
\texttt{source} & enum & Y & \{\texttt{rule}, \texttt{model}, \texttt{human}\}. \\
\texttt{confidence} & float & O & In [0,1]; tool-specific ranking score (not assumed calibrated). \\
\rowcolor{tblrow}
\texttt{review\_status} & enum & O & \{\texttt{unreviewed}, \texttt{approved}, \texttt{rejected}\}. \\
\texttt{reviewer\_id} & string & O & Pseudonymous reviewer id, if applicable. \\
\rowcolor{tblrow}
\texttt{layout\_zone} & enum/string & O & e.g., \texttt{body}, \texttt{header}, \texttt{footnote}, \texttt{caption}. \\
\texttt{note} & string & O & Free-form rationale/tag (e.g., ``hyphenation repair''). \\
\bottomrule
\end{tabularx}
\caption{Canonical span-edit provenance event schema used in this paper.}
\label{tab:app-schema-fields}
\end{table}

\paragraph{Versioning.}
We include \texttt{schema\_version} in each event. Minor versions may add optional fields without changing existing field meanings; major versions may change field meanings or constraints. All experiments in the current draft assume a fixed schema.

\subsection{Canonical JSONL and OCR operation examples}
\label{app:jsonl-examples}
We serialize events as JSONL (one event per line). This supports streaming pipelines and common NLP artifact formats while keeping events queryable and replayable.

\paragraph{Canonical example.}
The main text uses an illustrative correction \textit{Madifon}$\rightarrow$\textit{Madison} (model-assisted; confidence 0.74; no human approval). In our schema this is represented as:

\noindent\begin{tabularx}{\columnwidth}{@{}X@{}}
{\small\ttfamily
\{"schema\_version":"1.0.0",\allowbreak
"event\_id":"c2f1...",\allowbreak
"doc\_id":"doc\_017",\allowbreak
"page\_id":3,\allowbreak
"base\_revision":0,\allowbreak
"span\_start":1284,\allowbreak
"span\_end":1291,\allowbreak
"orig\_text":"Madifon",\allowbreak
"new\_text":"Madison",\allowbreak
"edit\_type":"substitute",\allowbreak
"source":"model",\allowbreak
"confidence":0.74,\allowbreak
"review\_status":"unreviewed",\allowbreak
"layout\_zone":"body"\}
}
\end{tabularx}

\paragraph{OCR-specific operations as span edits.}
We encode OCR cleanup as span replacements: \emph{hyphenation repair} (\texttt{"inter-\\national"}$\rightarrow$\texttt{"international"}) via \texttt{normalize} or \texttt{substitute} spanning the hyphen and break; \emph{line-break normalization} (newline$\rightarrow$space) via \texttt{normalize}; \emph{whitespace reflow} (collapse repeated whitespace / remove intra-word spaces) via \texttt{normalize}; \emph{split/merge boundaries} (e.g., \texttt{"NewYork"}$\rightarrow$\texttt{"New York"}) via \texttt{split/merge}; and \emph{paragraph merges/reflow} as a single larger-span replacement to avoid brittle micro-edits while preserving auditability.

\section{Provenance-Aware Filtering Method}
\label{app:filtering-method}

\subsection{Inputs and outputs}
\paragraph{Inputs.}
(i) Base text $T$ (raw OCR; \texttt{base\_revision=0}); (ii) a set of correction events $E$ in the schema of Table~\ref{tab:app-schema-fields}; and (iii) a trust policy $\pi$ over event metadata.

\paragraph{Outputs.}
A provenance-filtered variant $T_{\pi}$ constructed by replaying the subset of events selected by $\pi$, alongside an application trace (applied/skipped/conflicted events) for audit.

\subsection{Policy language and policies used in this paper}
A policy $\pi$ is a predicate on events. We use: \textbf{All corrections} $\pi(e)\equiv \texttt{true}$; \textbf{Confidence threshold} $\pi_{\tau}(e)\equiv (\texttt{confidence}(e)\ge \tau)$ with $\tau\in\{0.50,0.70,0.85\}$ (Table~\ref{tab:threshold-sensitivity}) and $\tau=0.70$ as the main provenance-filtered condition; and \textbf{Human-approved only} $\pi_{\text{approved}}(e)\equiv (\texttt{review\_status}(e)=\texttt{approved})$.

We emphasize that \texttt{confidence} is treated as a within-pipeline ranking signal, not a calibrated probability.

\subsection{Deterministic replay and overlap handling}
\paragraph{Ordering.}
Given selected events $E_{\pi}=\{e\in E:\pi(e)\}$, we sort by \texttt{span\_start} ascending, break ties by \texttt{event\_id}, and replay as base-anchored replacements.

\paragraph{Conflicts.}
Because all events are anchored to the same base revision, overlaps can be detected explicitly. When overlapping events are incompatible, we resolve deterministically under a fixed rule (e.g., prefer human over model over rule; prefer \texttt{review\_status=approved} over \texttt{unreviewed}; otherwise defer to adjudication). The key methodological point is that overlap handling is \emph{explicit and reportable}, rather than silently overwritten.

%% file: refs.bib
@inproceedings{smith2007tesseract,
  author    = {Smith, Ray},
  title     = {An Overview of the Tesseract OCR Engine},
  booktitle = {Proceedings of the Ninth International Conference on Document Analysis and Recognition (ICDAR 2007)},
  pages     = {629--633},
  year      = {2007},
  doi       = {10.1109/ICDAR.2007.4376991}
}

@article{nguyen2021postocrsurvey,
  author  = {Nguyen, Thi Tuyet Hai and Jatowt, Adam and Coustaty, Micka{\"e}l and Doucet, Antoine},
  title   = {Survey of Post-OCR Processing Approaches},
  journal = {ACM Computing Surveys},
  volume  = {54},
  number  = {6},
  pages   = {124:1--124:37},
  year    = {2021},
  doi     = {10.1145/3453476}
}

@inproceedings{neudecker2021ocreval,
  author    = {Neudecker, Clemens and Baierer, Konstantin and Gerber, Mike and Clausner, Christian and Antonacopoulos, Apostolos and Pletschacher, Stefan},
  title     = {A Survey of OCR Evaluation Tools and Metrics},
  booktitle = {Proceedings of the 6th International Workshop on Historical Document Imaging and Processing (HIP 2021)},
  pages     = {13--18},
  year      = {2021},
  doi       = {10.1145/3476887.3476888}
}

@inproceedings{evershed2014context,
  author    = {Evershed, John and Fitch, Kent},
  title     = {Correcting Noisy OCR: Context Beats Confusion},
  booktitle = {Proceedings of the First International Conference on Digital Access to Textual Cultural Heritage (DATeCH 2014)},
  pages     = {45--51},
  year      = {2014},
  doi       = {10.1145/2595188.2595200}
}

@inproceedings{nguyen2020bert,
  author    = {Nguyen, Thi-Tuyet-Hai and Jatowt, Adam and Nguyen, Nhu-Van and Coustaty, Micka{\"e}l and Doucet, Antoine},
  title     = {Neural Machine Translation with BERT for Post-OCR Error Detection and Correction},
  booktitle = {Proceedings of the ACM/IEEE Joint Conference on Digital Libraries (JCDL 2020)},
  pages     = {333--336},
  year      = {2020},
  doi       = {10.1145/3383583.3398605}
}

@article{lyu2021neural,
  author  = {Lyu, Lijun and Koutraki, Maria and Krickl, Martin and Fetahu, Besnik},
  title   = {Neural OCR Post-Hoc Correction of Historical Corpora},
  journal = {Transactions of the Association for Computational Linguistics},
  volume  = {9},
  pages   = {479--493},
  year    = {2021},
  doi     = {10.1162/tacl_a_00379}
}

@article{strange2014murder,
  author  = {Strange, Carolyn and McNamara, Daniel and Wodak, Josh and Wood, Ian},
  title   = {Mining for the Meanings of a Murder: The Impact of OCR Quality on the Use of Digitized Historical Newspapers},
  journal = {Digital Humanities Quarterly},
  volume  = {8},
  number  = {1},
  year    = {2014},
  url     = {http://www.digitalhumanities.org/dhq/vol/8/1/000168/000168.html}
}

@inproceedings{vanstrien2020impact,
  author    = {van Strien, Daniel and Beelen, Kaspar and Coll Ardanuy, Mariona and Hosseini, Kasra and McGillivray, Barbara and Colavizza, Giovanni},
  title     = {Assessing the Impact of OCR Quality on Downstream NLP Tasks},
  booktitle = {Proceedings of the 12th International Conference on Agents and Artificial Intelligence (ICAART 2020)},
  pages     = {484--496},
  year      = {2020},
  doi       = {10.5220/0009169004840496}
}

@inproceedings{hamdi2019analysis,
  author    = {Hamdi, Ahmed and Jean-Caurant, Axel and Sidere, Nicolas and Coustaty, Micka{\"e}l and Doucet, Antoine},
  title     = {An Analysis of the Performance of Named Entity Recognition over OCRed Documents},
  booktitle = {2019 ACM/IEEE Joint Conference on Digital Libraries (JCDL)},
  pages     = {333--334},
  year      = {2019},
  doi       = {10.1109/JCDL.2019.00058}
}

@inproceedings{boros2020alleviating,
  author    = {Boro{\c{s}}, Emanuela and Hamdi, Ahmed and Linhares Pontes, Elvys and Cabrera-Diego, Luis Adri{\'a}n and Moreno, Jose G. and Sidere, Nicolas and Doucet, Antoine},
  title     = {Alleviating Digitization Errors in Named Entity Recognition for Historical Documents},
  booktitle = {Proceedings of the 24th Conference on Computational Natural Language Learning (CoNLL)},
  pages     = {431--441},
  year      = {2020},
  doi       = {10.18653/v1/2020.conll-1.35},
  url       = {https://aclanthology.org/2020.conll-1.35/}
}

@article{ehrmann2024survey,
  author  = {Ehrmann, Maud and Hamdi, Ahmed and Linhares Pontes, Elvys and Romanello, Matteo and Doucet, Antoine},
  title   = {Named Entity Recognition and Classification in Historical Documents: A Survey},
  journal = {ACM Computing Surveys},
  volume  = {56},
  number  = {2},
  pages   = {27:1--27:47},
  year    = {2024},
  doi     = {10.1145/3604931}
}

@inproceedings{ehrmann2020clefhipe,
  author    = {Ehrmann, Maud and Romanello, Matteo and Fl{\"u}ckiger, Alex and Clematide, Simon},
  title     = {Overview of CLEF HIPE 2020: Named Entity Recognition and Linking on Historical Newspapers},
  booktitle = {Experimental IR Meets Multilinguality, Multimodality, and Interaction (CLEF 2020)},
  series    = {Lecture Notes in Computer Science},
  volume    = {12260},
  pages     = {288--310},
  year      = {2020},
  doi       = {10.1007/978-3-030-58219-7_21}
}

@inproceedings{ehrmann2022hipe,
  author    = {Ehrmann, Maud and Romanello, Matteo and Najem-Meyer, Sven and Doucet, Antoine and Clematide, Simon},
  title     = {Extended Overview of HIPE-2022: Named Entity Recognition and Linking in Multilingual Historical Documents},
  booktitle = {Working Notes of CLEF 2022},
  series    = {CEUR Workshop Proceedings},
  volume    = {3180},
  pages     = {1038--1063},
  year      = {2022},
  doi       = {10.5281/zenodo.6979577}
}

@inproceedings{ehrmann2020impresso,
  author    = {Ehrmann, Maud and Romanello, Matteo and Clematide, Simon and Str{\"o}bel, Phillip Benjamin and Barman, Rapha{\"e}l},
  title     = {Language Resources for Historical Newspapers: The Impresso Collection},
  booktitle = {Proceedings of the 12th Language Resources and Evaluation Conference (LREC 2020)},
  pages     = {958--968},
  year      = {2020},
  url       = {https://aclanthology.org/2020.lrec-1.121/}
}

@article{during2021impressoinspect,
  author  = {D{\"u}ring, Marten and Kalyakin, Roman and Bunout, Estelle and Guido, Daniele},
  title   = {Impresso Inspect and Compare: Visual Comparison of Semantically Enriched Historical Newspaper Articles},
  journal = {Information},
  volume  = {12},
  number  = {9},
  pages   = {348},
  year    = {2021},
  doi     = {10.3390/info12090348}
}

@article{during2024transparent,
  author  = {D{\"u}ring, Marten and Bunout, Estelle and Guido, Daniele},
  title   = {Transparent Generosity: Introducing the impresso Interface for the Exploration of Semantically Enriched Historical Newspapers},
  journal = {Historical Methods: A Journal of Quantitative and Interdisciplinary History},
  volume  = {57},
  number  = {1},
  pages   = {20--40},
  year    = {2024},
  doi     = {10.1080/01615440.2024.2344004}
}

@misc{moreau2013provdm,
  author       = {Moreau, Luc and Groth, Paul and others},
  title        = {PROV-DM: The PROV Data Model},
  howpublished = {W3C Recommendation},
  year         = {2013},
  url          = {https://www.w3.org/TR/prov-dm/}
}

@misc{lebo2013provo,
  author       = {Lebo, Timothy and Sahoo, Satya and McGuinness, Deborah and others},
  title        = {PROV-O: The PROV Ontology},
  howpublished = {W3C Recommendation},
  year         = {2013},
  url          = {https://www.w3.org/TR/prov-o/}
}

@inproceedings{ockeloen2013biographynet,
  author    = {Ockeloen, C. J. and Fokkens-Zwirello, A. S. and ter Braake, S. and Vossen, P. T. J. M. and de Boer, V. and Schreiber, A. T. and Legene, S.},
  title     = {BiographyNet: Managing Provenance at Multiple Levels and from Different Perspectives},
  booktitle = {Proceedings of the 3rd International Workshop on Linked Science (LISC 2013)},
  year      = {2013},
  url       = {https://linkedscience.org/wp-content/uploads/2013/04/paper7.pdf}
}

@article{romein2024htrprov,
  author  = {Romein, C. Annemieke and Hodel, Tobias and Gordijn, Femke and van Zundert, Joris and Chagu{\'e}, Alix and van Lange, Milan and Jensen, Helle Strandgaard and Stauder, Andy and Purcell, Jake and Terras, Melissa and van den Heuvel, Pauline and Keijzer, Carlijn and Rabus, Achim and Sitaram, Chantal and Bhatia, Aakriti and Depuydt, Katrien and Afolabi, Mary Aderonke and Anikina, Anastasiia and Bastianello, Elisa and others},
  title   = {Exploring Data Provenance in Handwritten Text Recognition Infrastructure: Sharing and Reusing Ground Truth Data, Referencing Models, and Acknowledging Contributions},
  journal = {Journal of Data Mining and Digital Humanities},
  year    = {2024},
  doi     = {10.46298/jdmdh.10403},
  url     = {https://jdmdh.episciences.org/10403}
}

@inproceedings{10.1145/3772363.3798570,
author = {Guo, Haoze},
title = {ConsentDiff at Scale: Longitudinal Audits of Web Privacy Policy Changes and UI Frictions},
year = {2026},
isbn = {9798400722813},
publisher = {Association for Computing Machinery},
address = {New York, NY, USA},
url = {https://doi.org/10.1145/3772363.3798570},
doi = {10.1145/3772363.3798570},
booktitle = {Proceedings of the Extended Abstracts of the 2026 CHI Conference on Human Factors in Computing Systems},
articleno = {185},
numpages = {5},
location = {
},
series = {CHI EA '26}
}

@inproceedings{10.1145/3774904.3792853,
author = {Guo, Haoze and Wei, Ziqi},
title = {Hidden-in-Plain-Text: A Benchmark for Social-Web Indirect Prompt Injection in RAG},
year = {2026},
isbn = {9798400723070},
publisher = {Association for Computing Machinery},
address = {New York, NY, USA},
url = {https://doi.org/10.1145/3774904.3792853},
doi = {10.1145/3774904.3792853},
booktitle = {Proceedings of the ACM Web Conference 2026},
pages = {8337–8340},
numpages = {4},
location = {United Arab Emirates},
series = {WWW '26}
}

@misc{guo2026temporaldriftprivacyrecall,
      title={Temporal Drift in Privacy Recall: Users Misremember From Verbatim Loss to Gist-Based Overexposure}, 
      author={Haoze Guo and Ziqi Wei},
      year={2026},
      eprint={2509.16962},
      archivePrefix={arXiv},
      primaryClass={cs.HC},
      url={https://arxiv.org/abs/2509.16962}, 
}

@misc{guo2026feedtaxonomyuserfacingcues,
      title={Behind the Feed: A Taxonomy of User-Facing Cues for Algorithmic Transparency in Social Media}, 
      author={Haoze Guo and Ziqi Wei},
      year={2026},
      eprint={2602.03121},
      archivePrefix={arXiv},
      primaryClass={cs.HC},
      url={https://arxiv.org/abs/2602.03121}, 
}

@article{ciccarese2013pav,
  author  = {Ciccarese, Paolo and Soiland-Reyes, Stian and Belhajjame, Khalid and Gray, Alasdair J. G. and Goble, Carole and Clark, Tim},
  title   = {PAV ontology: Provenance, Authoring and Versioning},
  journal = {Journal of Biomedical Semantics},
  year    = {2013}
}

@book{moreau2013introprov,
  author    = {Moreau, Luc and Groth, Paul},
  title     = {Provenance: An Introduction to PROV},
  series    = {Synthesis Lectures on the Semantic Web: Theory and Technology},
  volume    = {3},
  number    = {4},
  pages     = {1--131},
  publisher = {Morgan \& Claypool},
  year      = {2013},
  doi       = {10.2200/S00528ED1V01Y201308WBE007}
}

@inproceedings{pletschacher2010page,
  author    = {Pletschacher, Stefan and Antonacopoulos, Apostolos},
  title     = {The PAGE (Page Analysis and Ground-Truth Elements) Format Framework},
  booktitle = {Proceedings of the 20th International Conference on Pattern Recognition (ICPR)},
  year      = {2010},
  doi       = {10.1109/ICPR.2010.72}
}

@inproceedings{clausner2011scenario,
  author    = {Clausner, Christian and Pletschacher, Stefan and Antonacopoulos, Apostolos},
  title     = {Scenario Driven In-Depth Performance Evaluation of Document Layout Analysis Methods},
  booktitle = {Proceedings of the International Conference on Document Analysis and Recognition (ICDAR)},
  year      = {2011},
  doi       = {10.1109/ICDAR.2011.282}
}

@inproceedings{devlin2019bert,
  author    = {Devlin, Jacob and Chang, Ming-Wei and Lee, Kenton and Toutanova, Kristina},
  title     = {BERT: Pre-training of Deep Bidirectional Transformers for Language Understanding},
  booktitle = {Proceedings of NAACL-HLT},
  year      = {2019},
  doi       = {10.18653/v1/N19-1423}
}

@inproceedings{liu2019roberta,
  author    = {Liu, Yinhan and Ott, Myle and Goyal, Naman and Du, Jingfei and Joshi, Mandar and Chen, Danqi and Levy, Omer and Lewis, Mike and Zettlemoyer, Luke and Stoyanov, Veselin},
  title     = {RoBERTa: A Robustly Optimized BERT Pretraining Approach},
  booktitle = {arXiv},
  year      = {2019},
  eprint    = {1907.11692},
  archivePrefix = {arXiv}
}

@inproceedings{wolf2020transformers,
  author    = {Wolf, Thomas and Debut, Lysandre and Sanh, Victor and Chaumond, Julien and Delangue, Cl{\'e}ment and Moi, Anthony and Cistac, Pierric and Rault, Tim and Louf, R{\'e}mi and Funtowicz, Morgan and Brew, Jamie},
  title     = {Transformers: State-of-the-Art Natural Language Processing},
  booktitle = {Proceedings of EMNLP: System Demonstrations},
  year      = {2020},
  doi       = {10.18653/v1/2020.emnlp-demos.6}
}

@misc{tei_standoff,
  author       = {{Text Encoding Initiative}},
  title        = {The \texttt{standOff} Element (TEI P5 Guidelines)},
  howpublished = {TEI Guidelines reference},
  year         = {2026},
  url          = {https://tei-c.org/release/doc/tei-p5-doc/en/html/ref-standOff.html}
}

@misc{alto_loc,
  author       = {{Library of Congress}},
  title        = {ALTO: Technical Metadata for Layout and Text Objects},
  howpublished = {Standard (Library of Congress)},
  year         = {2022},
  url          = {https://www.loc.gov/standards/alto/}
}

@misc{hocr_spec,
  author       = {Bohnet, Klaus and others},
  title        = {hOCR: The Embedded OCR Workflow and Output Format (Specification)},
  howpublished = {Specification (GitHub repository)},
  year         = {2024},
  url          = {https://github.com/kba/hocr-spec}
}

@inproceedings{neudecker2019ocrd,
  author    = {Neudecker, Clemens and Baierer, Konstantin and Borgman, Don and Federbusch, Markus and Boenig, Matthias and Maier, Wolfgang and Reul, Christian and Puppe, Frank},
  title     = {OCR-D: An End-to-End Open Source OCR Framework for Historical Printed Documents},
  booktitle = {Proceedings of the 3rd International Conference on Digital Access to Textual Cultural Heritage (DATeCH 2019)},
  year      = {2019},
  doi       = {10.1145/3322905.3322917}
}

@misc{ocrd_mets,
  author       = {{OCR-D Consortium}},
  title        = {Requirements on Handling METS/PAGE},
  howpublished = {OCR-D Specification},
  year         = {2026},
  url          = {https://ocr-d.de/de/spec/mets}
}

@misc{ocrd_workflows,
  author       = {{OCR-D Consortium}},
  title        = {OCR-D Workflow Guide},
  howpublished = {OCR-D Documentation},
  year         = {2026},
  url          = {https://ocr-d.de/en/workflows}
}
